\documentclass[sn-chicago,author-year]{sn-jnl}

\usepackage{graphicx}%
\usepackage{multirow}%
\usepackage{amsmath,amssymb,amsfonts}%
\usepackage{amsthm}%
\usepackage{mathrsfs}%
\usepackage[title]{appendix}%
\usepackage{xcolor}%
\usepackage{textcomp}%
\usepackage{manyfoot}%
\usepackage{booktabs}%
\usepackage{algorithm}%
\usepackage{algorithmicx}%
\usepackage{algpseudocode}%
\usepackage{listings}%
\usepackage{graphicx}

\usepackage[caption=false,font=normalsize,labelfont=sf,textfont=sf]{subfig}

\usepackage{comment}

\theoremstyle{thmstyleone}%
\theoremstyle{thmstyletwo}%

\theoremstyle{thmstylethree}%

\begin{document}

\title[Article Title]{A Multimodal Conceptual Framework to Achieve Automated Software Evolution for Context-rich Intelligent Applications}


\author{\fnm{Songhui} \sur{Yue}}\email{syue@csuniv.edu}

\affil{\orgdiv{Computer Science Department}, \orgname{Charleston Southern University}, \orgaddress{\street{9200 University Blvd}, \city{North Charleston}, \postcode{29406}, \state{SC}, \country{United States}}}


\abstract{While AI is extensively transforming Software Engineering (SE) fields, SE is still in need of a framework to consider overall all phases to facilitate Automated Software Evolution (ASEv), particularly for intelligent applications that are context-rich instead of conquering each division independently. Its complexity comes from the intricacy of the intelligent applications, the heterogeneity of the data sources, and the constant changes in the context. This study proposes a conceptual framework for achieving automated software evolution, emphasizing the importance of multimodality learning. A Selective Sequential Scope Model (3S)  model is developed based on the conceptual framework, and it can be used to categorize existing and future research when it covers different SE phases and multimodal learning tasks. This research is a preliminary step toward the blueprint of a higher-level ASEv. The proposed conceptual framework can act as a practical guideline for practitioners to prepare themselves for diving into this area. Although the study is about intelligent applications, the framework and analysis methods may be adapted for other types of software as AI brings more intelligence into their life cycles. }




\maketitle

\section{Introduction}\label{sec1}
\begin{figure}[htbp]
	\centering
	\includegraphics[width=1\textwidth]{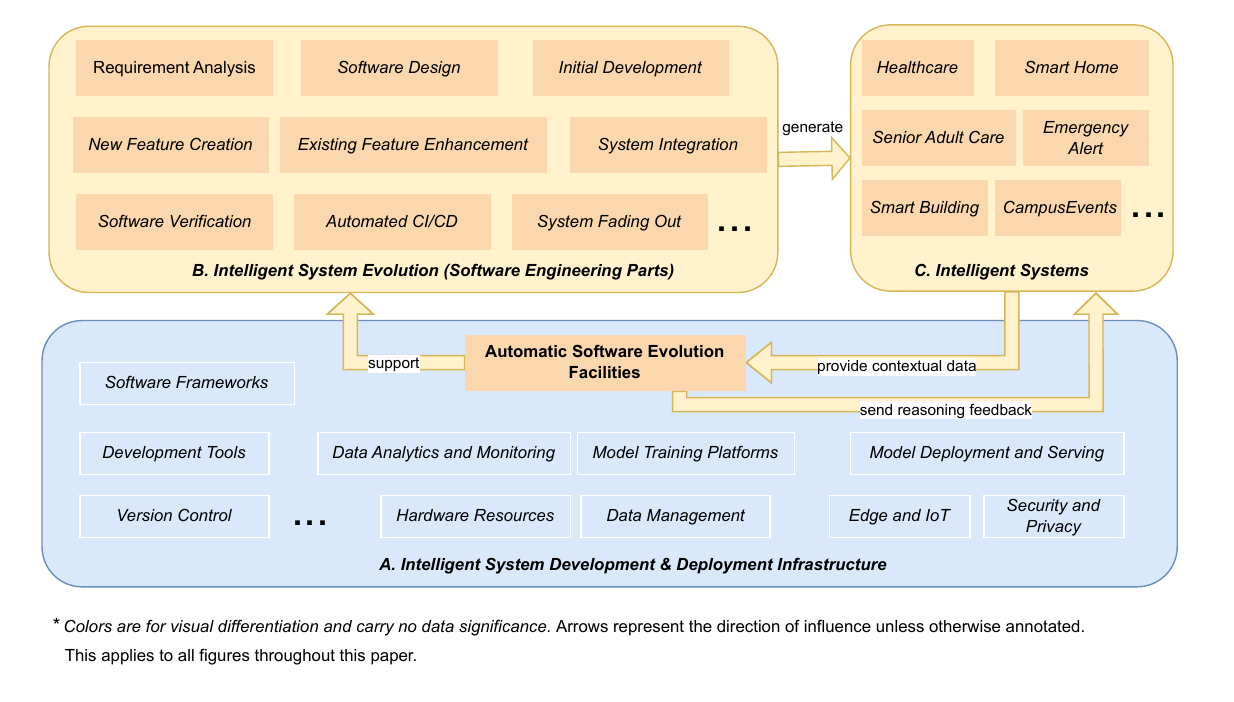}
	\vspace*{-5mm}
	\caption{Components of the Intelligent System Evolution Research}
	\label{position}
\end{figure}

With the advancements of Artificial Intelligence (AI)  and Natural Language Processing (NLP) in the past decades, especially the rise of the Large Language Model (LLM) and multimodality learning, software engineering fields welcome AI techniques to be employed in every aspect of software cycles. Meanwhile, the research of intelligent applications has continuously been a hotspot \citep{zhao2021understanding} because of the increasing amount of data of multimodalities generated in various domains. This type of software is designed to adapt to constantly changing scenarios of rich context \citep{zhao2021understanding, yue2021applying}, and some examples are listed in part C of figure \ref{position}. One primary characteristic of those applications is that a significant portion of their system behaviors is learned from continuous interaction with the users and environment involving detection and analysis of states and activities \citep{tzafestas2012intelligent, yang2013learning, cassavia2017discovering}, unlike applications of banking or insurance with more matured and stable business logic.

The rapid evolution of hardware and software wheels bring more capabilities to intelligent applications meanwhile making the creation and maintenance of that software more intricate  \citep{chu2021cloud, zheng2023intelligent}, both fields of software engineering and intelligent applications are eager for breakthroughs in higher-level automation (HLA) - collaboratively resolving the challenges by benefiting from AI techniques. One form of HLA of the two fields can be Automated Software Evolution / Evolvement (ASEv) \citep{o2001automated, ivers2020next}, which is a term derived from software evolution and Automated Software Engineering(ASE). Software evolution  \citep{2023_softwareevolution} refers to a continuous change from a lesser, simpler, or worse state to a higher or better state. ASEv can be the hero for Context-Rich Intelligent Applications (CRIA) in providing a fast response to the changing context, not only at the traditional context adaption (decision-making) level but also at the system evolution level in yielding new requirements and making relevant design and development automatically, as shown by the interaction arrows between the ASEv facilities, intelligent systems and the intelligent system evolution process in Figure \ref{position}.

For the purpose of introducing HLA or intelligence to software engineering, lots of correlated research has been applying Machine Learning (ML) and NLP techniques in requirement analysis \citep{cho2020comparative} and bug detection \citep{deshmukh2017towards} \citep{singh2020applying}, verification automation \citep{durelli2019machine} \citep{ma2022verification}, model-driven development \citep{wiesmayr2022supporting}. ASEv methodologies expect to build an automated procedure to make changes to an application or create a new application based on old systems statuses and available data if the current version of the system cannot fulfill the requirement of the present context.

One important essence of AI is fundamentally built upon data-driven mathematical analysis, logical reasoning, statistical learning, and algorithmic search. The data sometimes was referred to as the context of things. The things, such as applications, objects (e.g., camera, elevator), people, and even logic, include anything that has meaningful data that can benefit AI computing in making simple or complex context-aware decisions or recommendations, which can benefit from multimodality learning since the data are from various sources and may have different formats. Thus, ASEv for CRIA requires AI to thoroughly consider the context from both the software development side and the application running side. After the integration of various AI procedures, the final target is to produce some products that can be used by existing systems, methodologies, or human users. In software engineering areas, it can start by generating simple products like new code snippets.

However, the HLA of generation of new intelligent applications or features for CRIA requires researchers to have an overview of the combination between the ASEv process and the multimodality learning instead of only conquering partial of the task. A data-to-product multimodal conceptual framework proposed in this research, as a synthesis of interrelated components and variables, is designed as a response to this challenge, which can shine a light on related research. A conceptual framework is an integrated way of looking at problems \citep{liehr1999middle}, giving a broader understanding of a research problem, or telling a bigger map of possible relationships by joining together small individual concepts \citep{imenda2014there}. Intelligent applications are rich in contextual data, which is not only from environments but also from systems and software engineering processes. From this perspective, this framework is towards both "intelligent software" engineering and intelligent "software engineering," and the latter covers the applications that are context-rich in their software engineering process, which may not necessarily be a CRIA.

The subsequent sections of the article are organized as follows: Section 2 describes the background and related work. Section 3 outlines the key dimensions of the conceptual framework. Section 4 introduces a 3S model for categorizing solutions in the research of ASEv. Section 5 evaluates the framework and the 3S model by applying them to ASEv-related research. Section 6 discusses related topics of interest, limitations, and concerns. Finally, Section 7 provides the concluding remarks.

\section{Backgrounds and Related Work}\label{sec_backgrounds_related_work}
In order to better present the problem of achieving ASEv for CRIS, this section discusses the related concepts and literature in software evolution, data-to-product view, context dynamism, and multimodality.

\subsection{Automated Software Evolution}\label{subsec_software_evolution}
Software evolution refers to "a continuous change from a lesser, simpler, or worse state to a higher or better state" \citep{2023_softwareevolution}. To describe the HLA of CRIA, automated software evolution, a term derived from software evolution and Automated Software Engineering(ASE), is used to emphasize the updating and improving of the software. In this study, the term does not only include the process of software design, development, and maintenance; it also considers the deprecation of the software and the generation of the software from nowhere (generation of software based on patterns learned from other software or from the old, similar systems).

Historically, Computer Aided Software Engineering (CASE) \citep{vessey1995case} is a traditional approach that aids designers and developers in software engineering activities, including development and integration. However, CASE tools were criticized for being mainly descriptive. They provide neither a theoretical framework nor an attempt to explain the factors influencing adoption \cite{iivari1996case}. 

An interdisciplinary survey on challenges and state of the art in the evolution of automated production systems (APS) was conducted by Birgit et al. \citep{vogel2015evolution}. They argued that fundamental techniques such as variability modeling and tracing, which facilitate software evolution, are confined to the software domain. To develop those APS, not only software needs to be considered, but the system specifics and system design, which need to calculate the hardware, need to be considered throughout the whole process of product development. Although their research targets are more related to mechatronic automated products, such as washing machines or other automated systems that produce automated products, the perceived underlying rule can be applied to other types of software and hardware pairs. Take the iPhone's hardware ecosystem and software ecosystem as an example; not only do the operating system updates happen after hardware updates, but those mobile applications will also have to adapt themselves to the new features provided by new hardware and software. A vanward idea about hardware considers it as a new type of software \citep{10.1145/3102980.3103002}. Software often needs to be revised to meet changing market and regulatory demands.

Martin \citep{monperrus2018automatic} presented a survey on automatic software repair to resolve software bugs autonomously without interventions from human experts. The article discusses behavioral repair and state repair. Behavioral repair treats test suites, contracts, models, and crashing inputs as an oracle. State repair works at runtime using checkpoints and restarts, system reconfiguration, and invariant rehabilitation techniques. This survey work notably covers an extensive array of concepts in automatic software repair and spans a broad range of research areas, including not only software engineering but also operating systems, programming languages, and security.

A conceptual framework is presented by Nouredine et al. \citep{gasmallah2018developing} for modeling the classification of software architecture evolution approaches based on six explicit dimensions (What, Where, When, Who, Why, and How questions) treating the architectural viewpoint as a primary focus. Taking "When" as an example, it refers to the time of evolution, which consists of design-time evolution and runtime evolution. The former is at the design stages of software, allowing improvement and extension of the architecture by predicting the evolution earlier. The latter, as considered a primary theme in architecture adaptation, encompasses evolving at compile-time, load time, and dynamic time. As a comparison, this research studies a conceptual framework for resolving how multimodalities and AI shape the future automatic software evolution, which encloses architectural evolution. 

\subsection{Data-to-Product View}\label{subsec_data_to_product_view}
This subsection discusses a particular perspective of the "Data-to-Product" view, which describes a phenomenon of a software process that utilizes the data to produce the final product instead of utilizing the data to only generate intermediate decisions or recommendations. Compared with "requirement" and "feature" in terms like "data-to-feature" and "data-to-requirement," "product" is the final result of the software evolution and means high usability. So, it requires the AI techniques to be embedded into all ASEv life cycles, which will be mentioned in the framework and also illustrated in section \ref{sec_apply_model}.

\begin{figure}[htbp]
	\centering
	\includegraphics[width=0.95\textwidth]{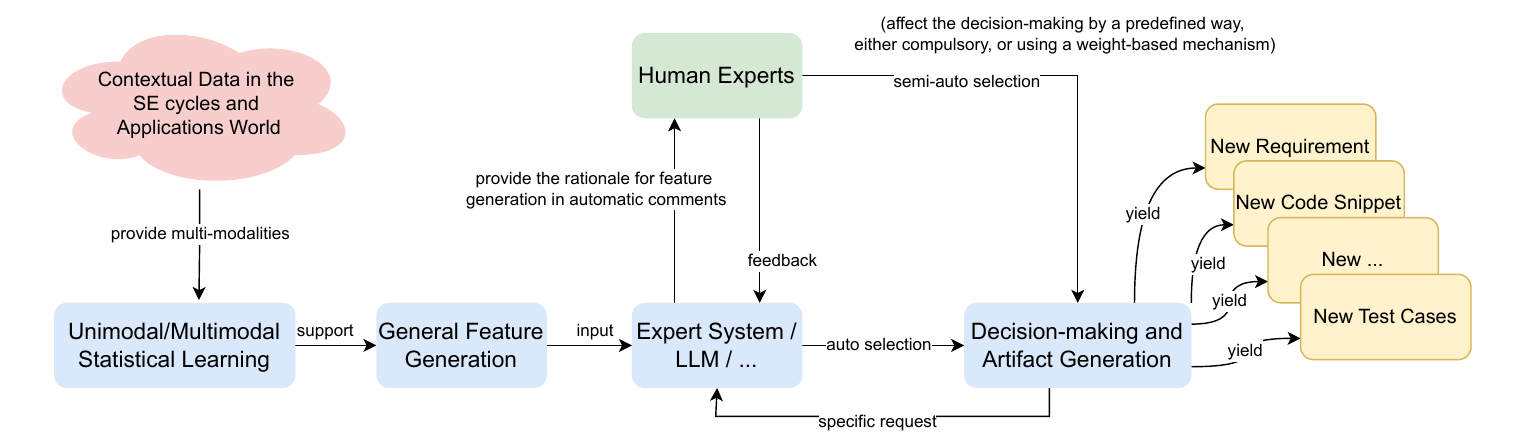}
	\vspace*{3mm}
	\caption{an auto / semi-auto software engineering decision-making process}
	\label{semi-auto}
\end{figure}

One example of Data-to-Product can be found in Fig. \ref{semi-auto}; some new requirements can be categorized by evaluating the general features generated by learning from contextual data \citep{talele2021software}. Auto selections or semi-auto selections can be made, and the selections will be sent to the decision-making and artifact generation step, where it can send a specific request based on combining extra considerations to form a loop of the decision-making process to improve the decision-making and artifact generation. The special request can ask for a new round of learning based on different features from a specific data set, feedback from other phases of product generation, or the result of the feature integration. This involves a process of optimization using a learning loop. Of course, the structure can be different. But this structure is just a demonstration that the product will be one that is final or approaching the final version. This view requires a constant collection of contextual data from SE cycles, applications, as well as environments.

More specifically, all types of data that can be used to improve the application or the development process can be counted as related context in Fig  \ref{semi-auto}. These data will be sent to an unimodal or multimodal learning model (more info can be found in Figure \ref{multimodal-architecture}). Some initial features can be learned from the data. After another round of learning through LLM or expert system, decisions can be made about the generation of products automatically or semiautomatically if human experts are needed in the decision-making process. The effect from human experts can be predefined, like using a weight-based mechanism. According to the source data, different types of products can be generated, such as new requirements, new code snippets, new test cases, etc. After combining those new products, an application can even be generated using an automated way of learning from source data of requirements, code artifacts, configuration, and the relationship data between them. The power of product generation and evolvement capability is unlimited, given enough raw data.

\subsection{White-Box, Black-Box, and Gray-Box Machine Learning Models}
Black-box or white-box are software engineering terms describing whether the inner logic is known to the examiners or not. ML models are classified using X-Box (white, black, or gray-box) w.r.t how to analyze, model, and encode the context data. As summarized by Michael et al. \citep{affenzeller2020white}: black-box ML techniques refer to methods that generate models where the internal workings are either concealed or too complex to be analyzed, producing outputs based on the inputs. In contrast, white-box modeling involves models with transparency that can be analyzed in detail. Decision tree models, linear regression models, bayesian networks, and fuzzy cognitive maps are the common examples of white-box models \citep{garcia2016grey, pintelas2020grey}; in comparison, deep neural networks, support vector machines, and Large Language Models (LLM) are the common examples of black-box models \citep{robnik2008explaining}.

Gray-box is the development of an ensemble of black and White-Box models in order to combine and acquire the benefits of both, building a more efficient global composite model \citep{bohlin2006practical, pintelas2020grey}. If an ensemble of ML algorithms contains both black and White-Box models, like neural networks and linear regression, it can be considered as a Grey-Box.

A most recent gray-box research is from Pan et al. \citep{pan2024unifying} and outlines a forward-looking roadmap for integrating Large Language Models (LLMs) and Knowledge Graphs (KGs). This roadmap features three primary frameworks: Firstly, KG-enhanced LLMs, which integrate KGs during both the pre-training and inference phases of LLMs to improve their understanding of acquired knowledge; Secondly, LLM-augmented KGs, which utilize LLMs to perform various KG tasks such as embedding, completion, construction, graph-to-text generation, and question answering; and thirdly, Synergized LLMs + KGs, where LLMs and KGs collaborate equally to enhance each other, facilitating bidirectional reasoning that leverages both data and knowledge.

Takeishi et al. \citep{takeishi2023deep} define deep gray-box models as compositions of data-driven models and theory-driven models, with the former being deep neural networks and the latter being human-understandable models with a theory as a backbone. They empirically analyze the sum of regularizers, including those that can prevent the theory-driven models from being ignored.

\subsection{Context Dynamism} \label{subsection_context_dynamism}
One primary characteristic of CRIA is that a great portion of their system behaviors is learned from continuous interaction with the users and environment involving detection and analysis of states and activities \citep{tzafestas2012intelligent, yang2013learning, cassavia2017discovering}, unlike applications of banking or insurance with more matured and stable business logic.

For CRIA, contextual data can change frequently, and so can the decision processes of the applications. Thus, contextual reasoning logic needs to adapt to the changes accordingly. The dynamism can sometimes lead to a new system feature \citep{yue2021applying}. For example, a smart elevator ships people in most scenarios, but for a period of time, it sees people convoying some large devices. Then, how it updates itself to fit the new normal and many more new normals could be a challenging problem. Some other types of dynamism can be perceived given enough data and time for analysis. Thus, the software team or a designated system should be smart enough to take care of updating the intelligent elevator system by identifying the dynamism automatically.

Thus, context dynamism refers to the dynamic aspects of entities, attributes, and states and the corresponding relationship changes within intelligent system domains \citep{yue2024csm}. One type of context dynamism can be the new joining of entity, attribute, or state, and an exemplified instance is "a new student is joining a smart campus." Other types include relationship changes, hotspot situation forming, and cycles and steps of transitions affecting decision-making \citep{yue2024csm}.

Although big changes for business-matured applications can be rare, some small or medium changes due to context changes are unavoidable, such as adaptation to regularities, new versions of the underlying software, or optimized workflow from analyzing contextual data. Although the main focus of ASEv in this study is CRIA, it can also be applied to business-stable applications. Changes in the data can result in changes in the code or configuration. A conceptual framework is needed to support the building of this automated application evolution process. 

\subsection{Multimodality}\label{subsec_multimodality}
Multimodality is extensively studied in AI / ML research \citep{massaro2012multimodal, baltruvsaitis2018multimodal, pena2023human} and industry engineering \citep{gaw2022multimodal, hou2024more}. It refers to a phenomenon of harnessing data from heterogeneous sources to achieve a better understanding of some target data or make a more precise decision by reasoning those data from multimodalities. Due to the intrinsic similarity, the popularity of multimodality can be derived from the data fusion \citep{dalla2015challenges, lahat2015multimodal}, while multimodality is a natural fact that exists in our daily life, as Tadas et al. pointed out in their article \citep{baltruvsaitis2018multimodal}: "Our experience of the world is multimodal - we see objects, hear sounds, feel the texture, smell odors, and taste flavors."

\begin{figure}[htbp]
	\centering
	\includegraphics[width=0.95\textwidth]{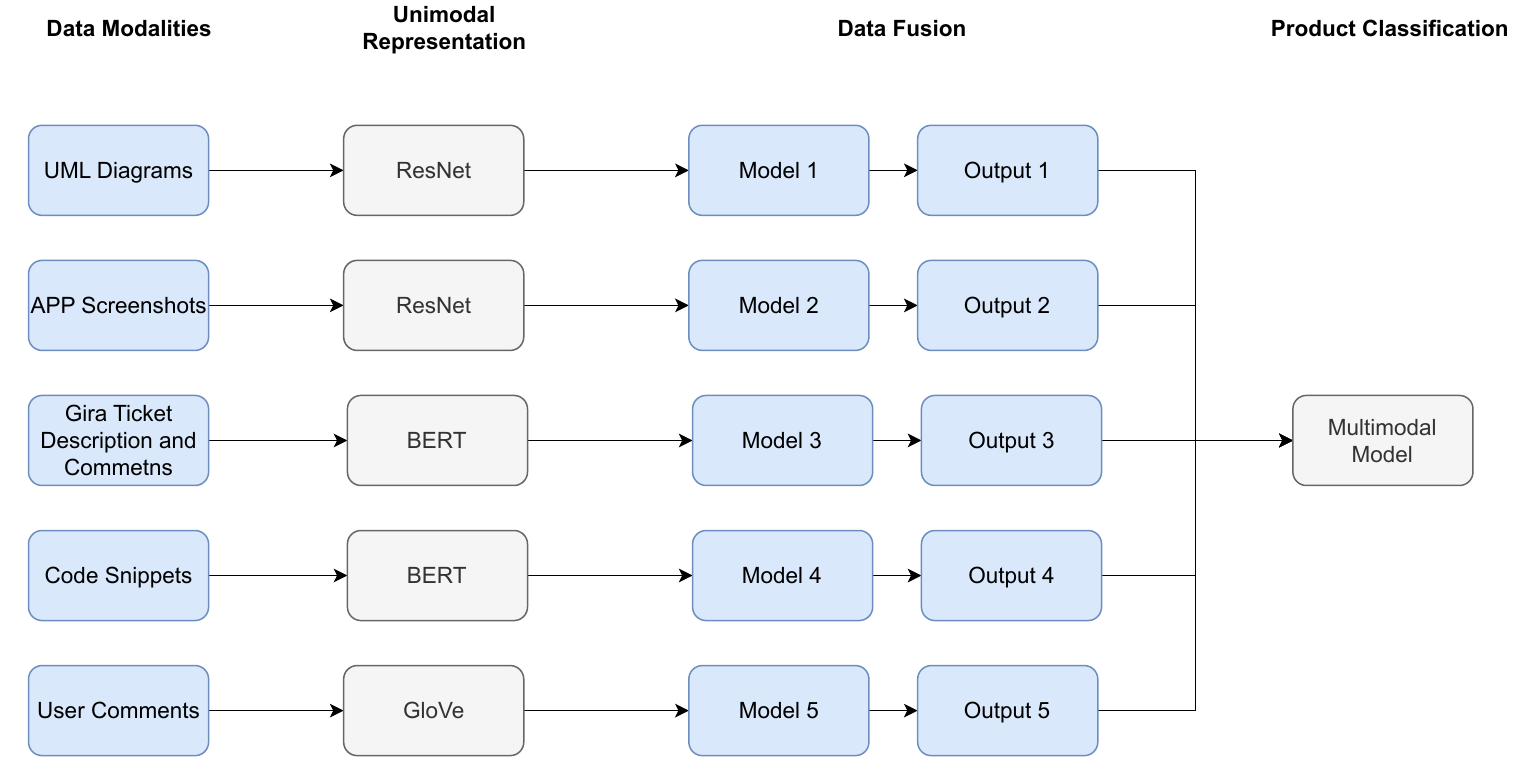}
	\vspace*{3mm}
	\caption{the Architectures of Multimodal Learning through Late Fusion within Software Engineering}
	\label{multimodal-architecture}
\end{figure}

Multimodal Deep Learning models typically consist of multiple neural networks, each specialized in analyzing a particular modality. The output of these networks is then combined using various fusion techniques, such as early fusion, late fusion, or hybrid fusion, to create a joint representation of the data.
A fascinating observation from empirical multimodal learning is that a model trained with multiple modalities can outperform a finely-tuned unimodal model, even on population data of the same unimodal task \citep{lu2024theory}.

A definition of multimodality for ML was also given by Letitia et al. \citep{parcalabescu2021multimodality}: A ML task is multimodal when inputs or outputs are represented differently or are composed of distinct types of atomic units of information. In this perspective, to achieve automated software evolution by utilizing data with different formats from various phases of the software engineering is a multimodal learning process, as shown in Figure \ref{multimodal-architecture}, which contains a structure adapted from the work \citep{pawlowski2023effective}. 

Thus, \textit{multimodal learning} in software engineering can be defined as \textit{the term to describe the learning process with data from heterogeneous sources of various phases of software engineering}. Data includes text, documents, code, images, videos, models, etc., from external and internal environments of applications. The definition emphasizes the differences between sources: as long as the sources are different, they contribute to \textit{multimodality}, regardless of whether their formats are the same or distinct.

\section{a Bigger Map and the Key Dimension/Feature Analysis}\label{sec_framework}
\begin{figure}[htbp]
	\centering
	\includegraphics[width=1\textwidth]{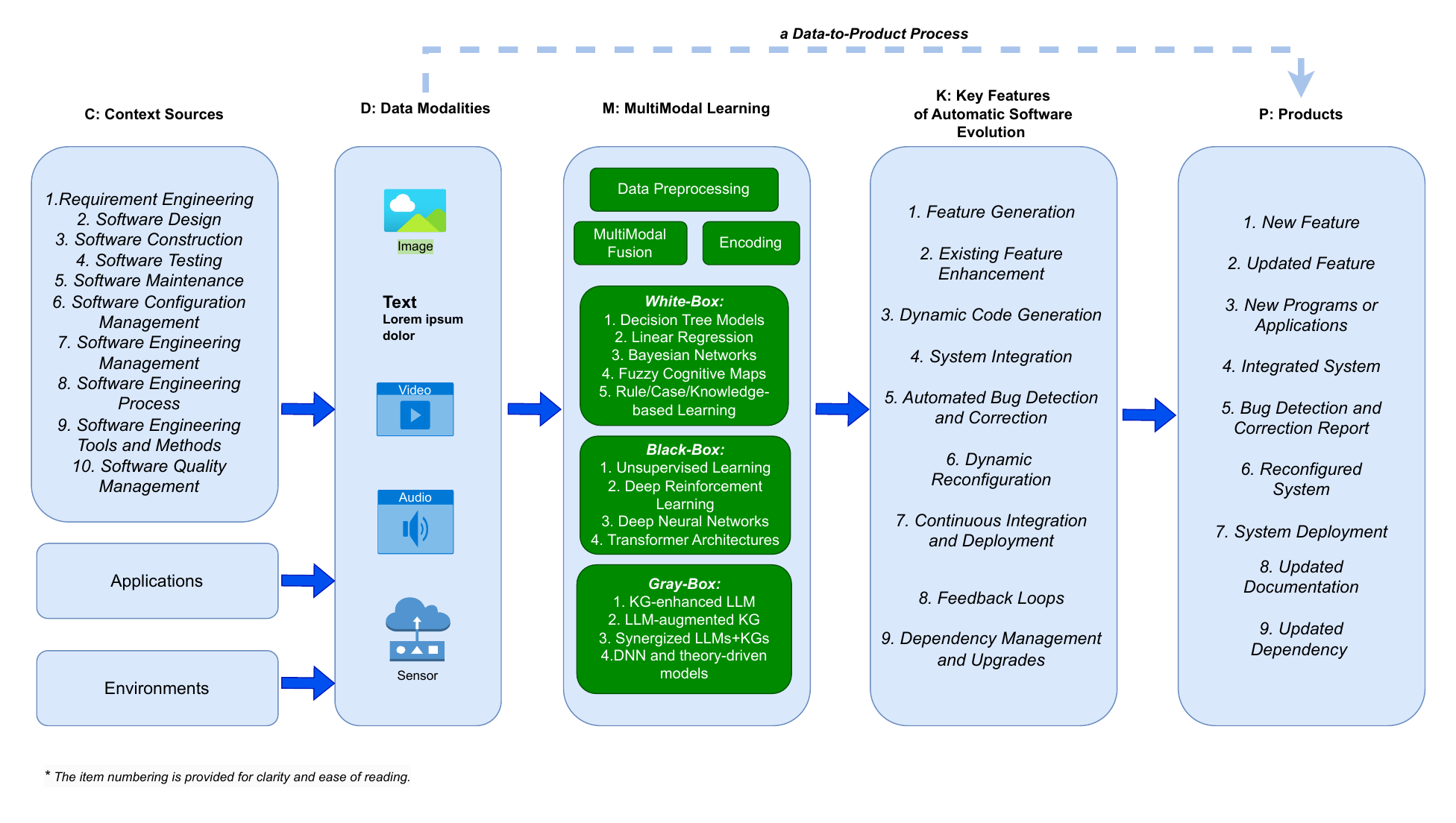}
	\vspace*{-5mm}
	\caption{a Data-to-Product Multimodal Conceptual Framework for ASEv}
	\label{framework}
\end{figure}

Figure \ref{framework} demonstrates a Data-to-Product multimodal framework for ASEv. It includes dimensions of Context Sources (C), Data Modalities (D), Multimodal Learning (M), Key features of ASEv (K), and Products (P). The internal relations among them are as follows: C provides data to D, D is the source of M, K is the key feature of ASEv, and M. P is the result of applying K. The process of generating P is not trivial and can need a tailored platform to achieve each of the tasks.

The items listed in the framework dimensions are not exhaustive, and they are examples that should be considered with higher priorities to the author's best knowledge. In the following subsections, key dimensions, exemplified items, and features of the framework are analyzed.

\subsection{Context Sources and Data Modalities}
Context sources for software design, development, and maintenance encompass various activities in any phase of the software engineering life cycle, as well as internal and historical data from intelligent applications and correlated environments. The underlying philosophy behind incorporating such a broad spectrum of context sources is that the more related details are collected, the better decisions can be made in generating the final software products. This assumption is based on a hypothesis that the learning methods can effectively leverage the diverse data pool. Contextual data derived from applications includes users' habits and other historical information, alongside sensor data from the surrounding environment, which can help to capture whether a user is correlated with a specific state, such as weather temperature, body movement, noise level, and more.

The software engineering process yields various types of data, including presentations, videos, discussion audio, images of graphical design, text from the requirement documents, and maintenance tickets, which record the steps of communications and complex business logic discussions. However, leveraging certain types of this data requires careful handling of privacy concerns and obtaining consent from participants \cite{saini2013privacy, kagan2023zooming}. Techniques have been developed to safeguard the privacy of individuals in videos, such as face obfuscation. A study of face obfuscation by Kaiyu et al. \cite{yang2022study} found that the features learned on obfuscated images are equally transferable when performing experiments with transfer learning to downstream tasks such as object recognition, scene recognition, and face attribute classification.

\subsection{Multimodal Learning}
The definitions of mulitmodalities are comprehensively studied by Letitia et al. \citep{parcalabescu2021multimodality}. Three types of definitions are discussed in their work, including human-centered (relies on human perceptual experience such as hearing and seeing), machine-centered (how is information represented), and ML task-related (inputs and outputs are represented differently or are composed of distinct types of atomic units of information). According to the ML task-related definition, it is multimodal learning even if only text data is involved and if the input and output formats are different or it is handling different units of data.

There are a bunch of techniques in multimodal learning. To better describe this component, a categorization method is particularly used by dividing them into white-box, black-box, and gray-box ML models. Firstly, they are terms that originated in the software engineering area. Secondly, the three categories can provide a complete coverage of the models. In each of the categories, specific methods are listed in the framework component, as shown in Figure \ref{framework}. Meanwhile, multimodal learning also considers how to preprocess data and the fusion of the data. So, this component in our framework consists of the subcomponents representing data processing tasks and machine learning methods that work with different inputs or outputs or distinct types of atomic units of information.

\subsection{Key Features and Products}
Key features of ASEv, as the objectives of multimodal learning, are essential components of the learning process. They are functional modules for building final products, including new features, programs, systems, and their updates. After all, the ultimate objective remains to be creating useful products. This perspective underscores a shift towards a Data-to-Product approach (the connection from D to P in Figure \ref{framework}), emphasizing the transformation of data into tangible outcomes rather than focusing solely on the evolution procedures. Consequently, the automated software evolution process must encompass a decision-making loop to address this aspect effectively, e.g., a new feature can be examined by a new phase of learning. Thus, the conceptual framework outlined here not only delves into the intricacies of various domains and subproblems but also explores the integration of them to fulfill the Data-to-Product view.

Feature generation and enhancement benefit the function designs and can be learned from the user's habits and changes in the context of the environment. They are mapped with the software requirement phase. A feedback loop module is particularly necessary for the reason of the second-order effects, as discussed in \citep{sutton2018reinforcement}, which can be observed due to the dynamic nature of human learning systems. When working with intelligent software agents, human users continually refine their models, beliefs, and expectations through trial and error, resulting in a learning process akin to reinforcement learning. In some cases, the learning dynamics may exhibit positive feedback loops, potentially amplifying smaller effects over extended periods.

Dynamic code generation deals with the software development phase. For example, the work \citep{li2022competition} delves into the realm of dynamic code generation, particularly focusing on competition-level code generation with AlphaCode. System integration, as an important consideration of today's complex system development, cannot be neglected in order to provide robust and fault-tolerant services. Automated bug detection and correction, as mapped to the phase of software testing, ensures the correctness and quality of the software. Configuration, deployment, and dependency management are other crucial factors that facilitate the automation of the overall software evolution process.

\section{3-S Model}\label{sec_3s_model}
\begin{figure}[htbp]
	\centering
	\includegraphics[width=0.75\textwidth]{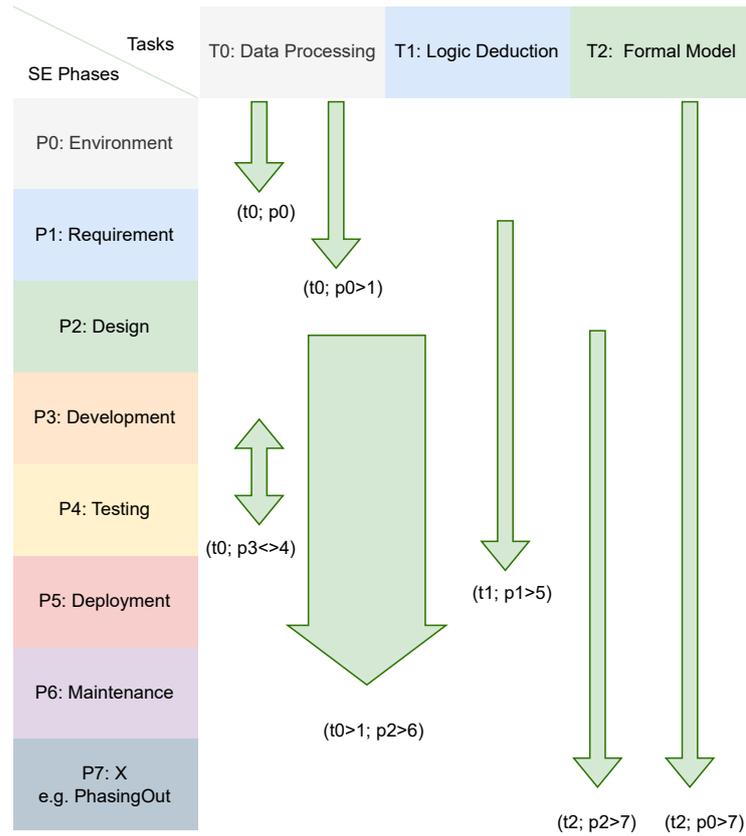}
	\vspace*{3mm}
	\caption{a Selective Sequential Scope Model (SSSM or 3S Model) for Smart (Intelligent) Software Engineering based on the software development life cycle and Multimodal Tasks}
	\label{3s}
\end{figure}

The tasks in achieving ASEv through multimodality learning correlate with the SE phases, as shown in the rows of Figure \ref{3s}. So do the products mentioned in the conceptual framework. One of the benefits of developing the conceptual framework is identifying products through analyzing the concepts related to ASEv and their relations. Each product is generated through an automation approach that can cover multiple phases of SE. Given some ASEv research working with generating a product mentioned in the ASEv conceptual framework, it is significant to have a way to clearly describe the research coverage in terms of the SE phases and tasks.

This section presents a Selective Sequential Scope Model (SSSM or 3S model) for categorizing research work that spans different scopes (one or more phases correlate with one or more tasks). It is sequential because it is practical to work on one phase based on its prior phase. As a counterexample, it is still not practical to directly connect the phases of environment and design without covering/considering the requirement phase. Although there could be ways to achieve that, which can happen in the future or beyond the author's knowledge, another model could be used to describe those approaches. This model is solely for the sequential coverage of the scopes.

\subsection{Tasks}\label{subsec_tasks}
In order to support the transition from data to product, it is necessary to consider another dimension, data-logic-model, as shown in Figure \ref{3s}. Data processing can include data modeling, preprocessing, fusion, extracting initial features, or high-level features. Some ML algorithms can be applied initially to learn some high-level features. Logics can include business logic or entity-relationship logic. The logic deduction is to use multimodality learning to generate the decisions, knowledge, or other types of models described using ML models. 

Logic deduction is the reasoning process, including utilizing LLM or other multimodality learning techniques. The result of multimodal learning is a model or system that can perform tasks or make predictions based on a more comprehensive and holistic understanding of the input data. For example, in a multimodality model that processes both textual and visual data, the product can have improved performance in tasks like image captioning \citep{yu2019multimodal}, sentiment analysis on multimedia content \citep{stappen2021multimodal}, or any other applications where combining information from different modalities provides a richer and more accurate representation of the underlying patterns or semantics.

The formal model is a way to represent those generated ML models or reasoning systems, or it is a visualization of them; either way, it can be used to represent the result of learning from the data. Generally speaking, models here refer to complex functions that represent the products or resolve the decision task given input data. Notably, general-purpose LLMs are not formal models in this dimension and are treated as a logic deduction tool if included in a methodology.   

\subsection{Software Engineering (SE) Phases}\label{subsec_se_phases_in_3s_model}
The main body of the SE phases is from the classical waterfall model. However, a special phase zero is added as the start phase, and phase seven is added as a flexible phase. Phase zero is the environment from which all the requirements and context data are based. Phase seven is named "PhasingOut" in Figure \ref{3s}. However, it can be documenting, configuration, or other things.

Particularly, the "phasing out" stage takes care of how to store or delete the data while considering privacy protection and other regulations. Meanwhile, the data processed in this stage may preserve the business logic and database logic so that it can be used for future development.

\subsection{Scope}
This 3S model is presented with two dimensions: tasks and SE phases. The scope of the 3S model means the intersections of the two dimensions covered by an ASEv research. The research can be analyzed concerning its relationship to the two dimensions in order to decide which scope it relates to. For example, if a study explores using UML design to generate the code and the test cases in an automated way, it can be viewed as covering p2 to p4 (including p3) in the SE phase dimension. If it is using data processing and logic deduction, it also covers t0 to t1 in the task dimension.

Some notations are defined to describe the scope in a simplified way. Firstly, a pair is used to represent the coverage in two dimensions. Secondly, \textgreater and \textless are used to describe the direction of the coverage. If both are used, it means that the study allows the affection of phases in two ways, e.g., P3 \textless\textgreater P4, the data model learned from the development phase can generate the test cases, and meanwhile, the model generated from the test phases can help with the code development. If there is no continuous coverage, then a comma is used to include indexes of phases or tasks that are not contiguous to each other. The arrows shown in Figure \ref{3s} are used to denote scopes. The type with a wider area covering multiple columns considers two or more tasks simultaneously. The type that covers only one column involves one task.

Here are some more examples: If the scope is represented by (t0, p0\textgreater 1), it covers one task and only spans over two nearby phases. Another common route can be (t0, p3\textgreater 4), which considers the code generation and test code or test cases generation according to the code. If the route is (t0, p3\textless4), then the research can be test-driven development.

\section{Apply the Framework and 3S Model to ASEv-related Research}\label{sec_apply_model}
This section presents some endeavors that target ASEv or can help with ASEv. The data-to-model conceptual framework is applied to analyze these studies, and the 3S model categorizes them. It is not an exhaustive search of the literature, and only some representative research is selected as a demonstration.

In a broad sense, deep learning falls under the category of machine learning, and LLMs are special deep learning models that are trained on large amounts of data. In this section, we differentiate them into two categories, machine learning and LLM-based methods, mainly due to LLMs being trained on significantly more extensive and more diverse data corpora.

\subsection{Agent-based Software Development}
ChatDev \citep{qian2023communicative}, a virtual chat-powered company for software development, brings together "software agents" from diverse social identities, including chief officers, professional programmers, test engineers, and art designers. "collaborative chatting" among those agents will take place when presented with a human client's request. The method of ChatDev enables the automatic crafting of comprehensive software solutions that encompass source codes, environment dependencies, and user manuals.

ChatDev employs the widely adopted waterfall model and divides the software development process into four distinct phases: designing, coding, testing, and documenting. Particularly, a chat chain is designed to facilitate breaking down each phase into atomic subtasks, allowing for proposing and validating solutions through context-aware communication.

Their work with agents depends on AI tools such as ChatGPT to generate code and evaluate the code across the stages of designing, coding, testing, and documenting. Since documenting is not listed in the 3S model, only the first three stages are considered. For adding phases such as documenting to the model, a customized extension can be made to the 3S model. Thus, according to the scope definition in the 3S model, the pair (t2, p2\textgreater4) can be used to represent the scope of this research. Similarly, the scopes of other research mentioned in this section are generated and shown in Table \ref{3s-categorization}.

\begin{sidewaystable}[htbp]
	\caption{Studies in ASEv Categorized by the 3S Model}
	\label{3s-categorization}
	\begin{tabular*}{\textwidth}{@{\extracolsep{\fill}} p{0.2\linewidth} p{0.2\linewidth} p{0.2\linewidth} p{0.2\linewidth}}
		\toprule
		\textbf{Methodologies} & \textbf{Studies} & \textbf{Modalities} & \textbf{Coverage} \\
		\midrule
		\multirow{1}{*}{Agent-Based}
		& ChatDev \citep{qian2023communicative} & chat transcript, LLM & (t2, p2\textgreater4) \\
		\midrule
		\multirow{1}{*}{Genetic Programming}
		& GenProg \citep{le2011genprog}  & souce code, test cases & (t0\textgreater1, p3\textless\textgreater4) \\
		\midrule
		\multirow{1}{*}{Ontology-based}
		& A lightweight semantic processing approach \cite{kaiya2005ontology}  & requirement specification, ontological elements & (t0\textgreater2, p1) \\
		\midrule
		\multirow{1}{*}{Case-based}
		& ACAI \citep{danilchenko2012automated}  & specification text, code template &  (t2, p3) \\
		\midrule
		\multirow{1}{*}{Stated-based}
		& CSMEngine \citep{yue2021applying}  & high level context data &  (t0\textgreater2, p0\textgreater2) \\
		\midrule
		\multirow{2}{*}{Machine learning}
		& DeepFix \citep{gupta2017deepfix} & C programs & (t0\textgreater2, p4) \\
		\cmidrule{2-4}
		& ArduCode \citep{canedo2020arducode} & source code, textual metadata, hardware configuration lists  &(t0\textgreater2, p1\&3)\\
		\midrule
		\multirow{3}{*}{LLM-based}
		& CodeBERT\citep{mashhadi2021applying} & source code & (t0\textgreater2, p4)\\
		\cmidrule{2-4}
		& Empirical studies on code generation using LLM-based tools\citep{nejjar2023llms, liu2024empirical} & prompts, modalities of LLMs  &(t1, p1\&3)\\
		\cmidrule{2-4}
		& AID-an LLM-powered test case generation framework \citep{liu2024llm} & prompts, modalities of LLMs  &(t1, p4)\\
		\bottomrule
	\end{tabular*}
\end{sidewaystable}

\subsection{Genetic Programming}
GenProg \citep{le2011genprog} uses an extended form of genetic programming (GP) to evolve a program variant to automatically repair it. GP uses computational analogs of biological mutation and crossover to generate new program variants, which are evaluated by a user-defined fitness function; GenProg uses the input test cases to evaluate the fitness, and individuals with high fitness are selected for continued evolution. This GP process is successful when it produces a variant that passes all tests and can be encoded with required behaviors and bugs. In particular, GenProg uses only statements from the program itself to repair errors and does not invent new code.

\subsection{Ontology-based}
A lightweight semantic processing approach \cite{kaiya2005ontology} was designed to develop requirement documents using an ontology technique, where they establish a mapping between a requirement specification and ontological elements. Changes in a requirement ontology can be used to predict the next changes, which enables the method to improve the completeness of requirements specification incrementally. They are working only with the requirement phase. However, their method involves all three tasks.

\subsection{Case-based}
An Automated coding system - ACAI (Automated Coder using Artificial Intelligence) is proposed by Danilchenko et al. \citep{danilchenko2012automated}. They treat code generation as a design problem, and the solutions are plans. Thus, ACAI is built with combining the techniques of routine design, template-based programming, and case-based reasoning. By accepting the objective and the specifications for how to achieve the objective, ACAI yields the result in the form of a working Java program. ACAI highly relies on tailored input templates and program templates, and more importantly, the user of ACAI should design the target code by applying the mechanisms the system provides. Thus, this system is categorized as working on the formal model in the phase of development without covering automated data processing and logic deduction.

\subsection{State-based}
CSMEngine \citep{yue2021applying} is proposed to automatically model and reason high-level context. Their study applied the methodology to an intelligent elevator system. CSMEngine is based on context state machine modeling methods \citep{yue2017state} and CSM-H-R \citep{yue2024csm} is a framework as an extension of the core model, to facilitate the modeling of the context dynamism. CSMEngine works with high-level context information, which is the data fused from the environment or applications. After automatically modeling the collected context, it can reason the context to identify new rules the applications can adapt to or the design process can utilize.

\subsection{Machine learning}
Fixing bugs is an essential activity in the maintenance of software. Bugs can be identified in the testing phase or reported by analyzing the failure information of a running system. No matter how the bugs are identified, they are supposed to be corrected. In this study, bug fixing is considered to be in the development and testing phases of the waterfall model.

DeepFix \citep{gupta2017deepfix} is a method for fixing common C language errors by deep learning. They treat the problem of fixing a programming error as a sequence-to-sequence learning problem, which requires a program to be represented as a sequence. They constructed the dataset by collecting correct programs and erroneous programs. Through the mutation of correct programs, they can build training examples that consist of the mutated programs and the fixes. Their trained model achieved a successful fix rate of 27\%. By collecting data from code examples and designing the deep learning neural network architecture for an attention-based sequence-to-sequence model, long-term dependencies can be captured to utilize the context in fixing the bugs.

ArduCode \citep{canedo2020arducode} is a predictive framework for assisting automation engineers in classifying automation code, finding similar code snippets, and reasoning about the hardware selection of sensors and actuators to resolve the requirement challenges from hardware and software integration. They use paragraph embedding in classification and achieve precision that is close to human annotation. Hardware recommendations are studied using autoencoder models. Thus, their work is categorized in the 3S model as working with the requirement and development.

\subsection{LLM-based}
CodeBERT\citep{feng2020codebert} is a transformer-based neural model that has been pre-trained on an extensive corpus of source code. Ehsan et al. \citep{mashhadi2021applying} apply it for automated program repair of Java simple bugs. Through experiments, they found the method can generate fix codes in 19-72\% of the cases with different types of their datasets, which are exactly the same as the fix codes implemented by developers.

Nejjar et al.  \citep{nejjar2023llms} conducted an empirical study on code generation tasks using LLM-based tools; specifically, the code generation tasks include a matrix multiplication function, scripts for data analysis, and data visualization. They performed the study by providing prompts to the tools, including GPT3.5, GPT4, Bing Chat, Google Bard, and YouChat. Their report shows the quality of code w.r.t. correctness, efficiency, comprehensibility, and overall rating. Liu et al. \citep{liu2024empirical} conducted an empirical study on generating software code for safety-critical domains using GPT-4. Compared with Nejjar's work, this study applied more approaches for utilizing the LLM tool, including overall requirements, specific requirements, and augmented prompts. The approach they applied is working on requirements and code development. The LLMs they used are general-purpose models, not formal models specifically for achieving ASEv tasks. They use LLMs only as a logic deduction tool. 

Liu et al.  \citep{liu2024llm} explore an LLM-powered test case generation method. A framework called AID is proposed. The framework consists of three main stages: the generation of program variants, the generation of program inputs, and differential testing. AID first feeds problem descriptions, program descriptions, and programs under test into LLMs to generate program variants and test inputs; then a diversity-first differential testing is applied to create test cases. AID is considered to be working in the phase of testing and the task of logic deduction.

\subsection{Summary}
The research presented in this section is not based on a comprehensive survey. However, to the best of the authors' knowledge, little research covers more than two stages in the SE phase dimension. As more data is collected and multimodal learning is applied to more specific tasks, it is predicted that future research will cover a broader scope so that a higher level of automation of the software evolution process can be achieved.

\section{Discussion}
This section presents an anticipated future trend and related prospective exploration directions and discusses some interesting topics that are closely relevant to the conceptual framework but were not discussed thoroughly in the main body due to the article's focus on multimodality and data-to-product process. Open Research Questions (ORQ) are specially identified for advancing the work of the proposed ASEv conceptual framework and the 3S model.

\subsection{Future Trends}
A trend is anticipated in research regarding leveraging data across multiple software engineering phases. Several prospective avenues for exploration are envisioned as follows:

\begin{enumerate}
	\item Initiating tasks, such as code generation, from raw environmental data or other types of contextual data.
	\item Implementing logic adaptation strategies to dynamically adjust software behavior based on evolving requirements.
	\item Harnessing Artificial General Intelligence \citep{kumpulainen2022artificial} to map and synthesize code logic, followed by iterative optimization and refinement processes.
\end{enumerate}

\subsection{Full Automation and Semi-automation}
All the aforementioned ASEv features could be approached through two methods: full automation (with minimum exterior intervention, e.g., initial configuration) and semi-automation (with significant exterior assistance, e.g., needing help in data generation, logic deduction). When simple tasks can be initiated through automated means, complex tasks necessitate an initial semi-automated approach, with the potential to transition gradually to full automation, for the initial stages of achieving full automation may pose technical challenges. This study presents a conceptual framework to facilitate the full automation of software evolution. Identifying tasks suitable for full automation and those requiring a semi-automated approach can help to apply a gradual progression toward automation as data thresholds are met and accumulated.

\subsection{X-Box Learning}
Explainability does not come for free. In order to make the model and learning process more understandable, extra efforts should be made to combine white-box learning with black-box learning to represent the knowledge in a way that human beings can comprehend. Although black-box learning models may be more efficient in time when dealing with large datasets and more accurate, as shown in many studies in the past decades, it does not mean they are always more accurate. The work of Emmanuel et al. \citep{pintelas2020grey} suggests white-box models or gray-box models can outperform black-box models, especially when the data amount is small. So, the real performance of various ASEv tasks involved needs to be investigated in practice.

\subsection{Practical Managerial Significance}
Management is a crucial facet in software engineering practices and theories \citep{gilb1988principles, karolak1995software, tsui2022essentials}, including project management, people management, progress management, quality management, risk management, and so on. The 'manager side' can be human participants and management tools, while the 'managed side' consists of human participants, software, and other control points. Practical Managerial Significance (PMS) can be a significant concern, as the ASEv process can affect the human participants and the software on both sides. Although this article primarily focuses on technical aspects and thus does not delve deeply into management concepts within the framework or the 3S model, exploring the integration of management practices in the framework and discussing the ASEv's PMS remains an interesting direction for future research.

\subsection{Data Amount and Privacy}
The prerequisites of multimodality learning rely on the availability of sufficient data; if the data does not meet a threshold, effective training sometimes becomes impossible. Privacy considerations often present obstacles to data acquisition, and the concerns should be thoroughly discussed and resolved. For some types of intelligent systems, if they are for public usage and have few privacy concerns, obtaining data from those intelligent systems' context could be more accessible. Nevertheless, it might still be hard to get the data from the software development cycles to protect the privacy or other ethical concerns of the software development participants. Possible strategies can be leveraging data from different software types and conducting cross-domain research to mitigate data scarcity within specific software development cycles.

\subsection{Other Ethical Concerns}
There is a term in sociology called Technological Determinism (TD) \citep{hallstrom2022embodying}, which suggests that technology is not just a tool for human users to achieve goals, but rather it controls and shapes human behaviors. This subsection is not to support or discuss TD, but among many influences that technologies can bring to human society, persuasion can particularly be an ethical concern of ASEv and the framework due to the applications of black-box models such as LLMs, which exhibit the traits of persuasion \citep{carrasco2024large} and the possibility of being persuaded \citep{zeng2024johnny}. In the analysis of the interaction between intelligent software agents and human users \citep{burr2018analysis}, technologies to influence human user behavior towards specific objectives are examined, such as increasing the time spent on a website. This goal-driven behavior has been previously discussed in the literature \citep{cristianini2010we}. It is imperative to critically analyze and regulate any technology with the potential to steer human behavior in the processes and products of ASEv.

\subsection{Limitations and Open Research Questions}
This conceptual framework considers the software engineering phases as the context sources and as one dimension in the 3S model. However, it's important to note that different methodologies of a specific phase can influence the workflow in a nuanced way, a factor that needs to be carefully considered. For instance, as a crucial consideration in the design phase and from the perspective of comparing technical-driven and tailored-usability-driven, system design methodologies can include hard systems, soft systems \citep{checkland2020soft}, and socio-technical design \citep{baxter2011socio} as a relatively balanced approach. Soft systems take into account human activity dynamics and their tailored needs \citep{burge2015overview}, while hard systems focus more on techniques and engineering methods. Such distinctions can lead to different data or logic flows under the framework of ASEv. Thus, the below ORQs are proposed for the framework and the 3S model.

\begin{enumerate}
	\item Extend the framework by considering various methodologies of SE phases. For example, in the design phase, the system design methodologies can be emphasized for a particular type of software. 
	\item The framework is a general one that treats software as a whole. Software in different domains may have varying needs in the process of ASEv, so another dimension to consider is the specific domains. Besides, domains can be a third axis for the 3S model. 
	\item Expand the framework to incorporate ethical concerns and social impacts, such as PMS, providing valuable insights for software management.
\end{enumerate}

\section{Conclusion}\label{sec_conclusion}
ASEv of CRIS is imperative due to the unceasing accumulation of contextual data from multiple sources and the intrinsic complexity and dynamism of the business logic of those systems. The constantly changing context can bring intelligent applications with new and updated requirements, which may be beyond the capability of manual design and development processes by system analysts, designers, and developers. In the meantime, multimodality learning has shown its advantages in utilizing the data to resolve real-world problems and is promising in helping achieve ASEv.

This article studies the blueprint using multimodality learning by crafting a data-to-product conceptual framework for ASEv. In order to categorize the research in this area, a 3S model is developed, using scope to represent the coverage of each research. The generated model can be an efficient method to categorize the related research when making the comparisons, as demonstrated in this study. This research is a preliminary step towards automated software evolution, and the proposed framework can act as a practice guideline for practitioners to prepare themselves for diving into this research area. Although the study is motivated by CRIA, the framework and analysis methods may be adapted for other types of software as AI brings more intelligence into their life cycles.

\backmatter

\bmhead{Acknowledgments}

The author claims that there is no conflict of interest.





\bibliography{references}

\end{document}